\documentclass{revtex4}

\usepackage{graphicx}
\usepackage{epsfig}
\begin{document}
\bibliographystyle{revtex}


\title{Young Physicists' Forum}



\author{T.~Adams}
\noaffiliation
\author{M.~Bishai}
\noaffiliation
\author{K.~Bloom}
\email[Contact author: ]{kenbloom@umich.edu}
\altaffiliation{YPF Steering Committee}
\noaffiliation

\author{V.~Boisvert}
\altaffiliation{YPF Steering Committee}
\noaffiliation
\author{L.~Coney}
\noaffiliation
\author{R.~Erbacher}
\altaffiliation{YPF Steering Committee}
\author{B.T.~Fleming}
\altaffiliation{YPF Steering Committee}
\author{J.~Formaggio}
\noaffiliation
\author{D.~Gerdes}
\noaffiliation
\author{A.~Green}
\noaffiliation
\author{S.~Heinemeyer}
\altaffiliation{YPF Steering Committee}
\author{K.~Hoffman}
\altaffiliation{YPF Steering Committee}
\noaffiliation
\author{B.~King}
\noaffiliation
\author{J.~Krane}
\altaffiliation{YPF Steering Committee}
\noaffiliation
\author{S.~Lammers}
\noaffiliation
\author{K.~Lynch}
\noaffiliation
\author{D.~Marfatia}
\altaffiliation{YPF Steering Committee}
\author{J.~McDonald}
\altaffiliation{YPF Steering Committee}
\noaffiliation
\author{K.~McFarland}
\noaffiliation
\author{G.~Moortgat-Pick}
\altaffiliation{YPF Steering Committee}
\noaffiliation
\author{T.~Nunnemann}
\noaffiliation
\author{M.~Palmer}
\altaffiliation{YPF Steering Committee}
\noaffiliation
\author{M.~Popovic}
\noaffiliation
\author{C.~Potter}
\altaffiliation{YPF Steering Committee}
\author{A.~Soffer}
\altaffiliation{YPF Steering Committee}
\noaffiliation
\author{Z.~Sullivan}
\altaffiliation{YPF Steering Committee}
\author{M.~Toharia}
\altaffiliation{YPF Steering Committee}
\noaffiliation
\author{W.~Walkowiak}
\noaffiliation
\author{G.~Zeller}
\altaffiliation{YPF Steering Committee}
\noaffiliation


\date{\today}

\begin{abstract}
The Young Physicists' Forum was an opportunity for the younger members
of the particle-physics community to gather at Snowmass 2001 and to
study and debate major issues that face the field over the next twenty
years.  Discussions were organized around three major topics: outreach
and education, the impact of globalization, and building a robust and
balanced field.  We report on the results of these discussions, as
presented on July 17, 2001.
\end{abstract}

\maketitle



While the field of high-energy particle physics (HEP) has a clear plan
for the coming decade, the period beyond that is uncharted territory.
Based on existing measurements and theories, we expect that there are
exciting new phenomena that could be discovered very soon -- new
particles that may be observed at the Tevatron or the Large Hadron
Collider (LHC), $CP$ violation in the $b$-quark system observed at $B$
factories, and neutrino mixing seen in both accelerator and
non-accelerator experiments.  These and other topics make for a very
rich program of study in the coming years.  But the next generation of
facilities and experiments that will be needed to explore these new
phenomena are technically challenging, and are expected to cost
billions of dollars.  They will require a large number of accelerator
and experimental particle physicists to build and operate them, and a
strong community of theoreticians to help interpret the results and
suggest new lines of exploration.  In addition, these facilities will
be built in a new social and political climate.  It is unlikely that
any single country can build and operate any such facility alone, so a
greater level of cooperation will be required across national and
continental boundaries.  Financial resources are limited, so difficult
decisions will have to be made about which facilities are built, and
how to set priorities for their use.  Finally, governments and the
general public will have to be convinced that such expensive
undertakings are worthwhile and in their interest.

It is the young physicists of today -- students, postdocs, and junior
faculty -- who will be leading these efforts in the long-term future,
and it will be their task to advance the field of particle physics in
this new environment.  Decisions that are taken now could affect this
generation for the duration of their careers.  To give the younger
members of the community a voice in these decisions, the Snowmass 2001
Organizing Committee scheduled a ``Young Physicists' Forum'' as part
of the workshop.  Here, we report on the studies and discussions that
were held for that event~\cite{ref:YPF}.  More than 200 young
physicists participated in the Snowmass workshop, and many were
engaged in the activities of the Forum.

The Forum was organized by a broad-based committee of students and
postdocs, chaired by K. Bloom.  The committee included theoretical,
experimental, and machine physicists, working in a variety of
subfields, with representatives from the United States and abroad.
After some initial discussions, working groups were formed to explore
three major topics.  These were:
\begin{enumerate}
\item Outreach and Education -- How can we communicate the excitement
of the field to the general public and influential groups?  How can we
develop their support so that particle physics will be a robust and
growing field in the long-term future?
\item Globalization -- How will the field operate in a more global
and diverse setting in the future?  What new organizational structures
are needed?  How does the U.S. maintain a strong program in this
context?
\item Balancing and Building the Field -- What kind of physics and
facilities do we want in the next twenty years, and what choices must
we make to get them?  What disciplines do we need to promote to
achieve these goals?  Where will these facilities be?  How do we
attract and keep people in the field so that we maximize the use of
new facilities?
\end{enumerate}
The working groups, led by J. McDonald, J. Krane, and Z. Sullivan, did
background research and prepared documentation on their topics in
advance of Snowmass.  This work, along with an introductory talk by
High-Energy Physics Advisory Panel (HEPAP) long-range-planning
subpanel member K. McFarland, was presented at an introductory meeting
for young physicists on July 3, which was attended by about 100
people.  There, the working groups expanded, and prepared for a
``Vermont-style,'' young-physicists-only town meeting that was held on
July 10.  In contrast to typical HEP town meetings, with a scripted
speaker list, no particular topical focus, and no resolution of any
issues, the 100 participants at this meeting had focused discussions
on the three major topics, guided by questions that the working groups
had prepared.  In many cases, straw-poll votes were taken to gauge
whether there was any consensus of opinion on each topic.  Meanwhile,
throughout the Snowmass workshop, the Young Physicists Panel
(YPP)~\cite{ref:yppweb} was conducting a wide-ranging
survey~\cite{ref:ypppoll} of the entire community on similar topics,
reaching over 1500 physicists.  The results of all of these studies,
surveys and discussions were presented to the Snowmass workshop on
July 17, in an evening session that attracted about 200 attendees,
young and not-as-young, who engaged in constructive debate on these
topics.

The following sections explore each of the three major topics,
presenting our research and outcomes of various meetings.  

\section{Outreach and Education}
\subsection{Introduction}
There is general agreement in our field that efforts to educate the
public about particle physics, usually called ``outreach,'' are
worthwhile and important for the long-term future of the field. The
American public is more likely to support our research if they
understand what we are trying to do and what makes it interesting.  We
participate in particle-physics research because we are excited about
understanding the fundamental nature of matter, space, and time, and a
properly informed public should be excited too.  We have an excellent
product, but we have to start and maintain a strong effort to sell it.

This outreach effort, like everything else, requires money and people.
Does our allocation of resources to outreach match our needs?  Here,
we review outreach activities taking place at various institutions,
report the results of discussions about outreach at Snowmass, and
present a plan of action for future activities.

\subsection{Background}
There are many existing outreach activities in HEP that build the
foundation for a greater involvement of young physicists. These
activities include outreach to the general public; outreach to
Congress, the funding agencies and the media; and teaching and
mentoring.  This outreach work is usually not an integral part of our
research programs, and is therefore contingent upon the volunteer
efforts of especially motivated people, and sometimes upon special
sources of funding.

Outreach to the general public has many aspects, all of which cannot
possibly be enumerated here.  We highlight some of the higher-profile
activities at national laboratories and universities. University open
houses and lab tours provide great opportunities for public outreach.
They enable the public to see first-hand our facilities, our
scientific accomplishments, and the people behind the scenes. These
activities reach several tens of thousands of people a year.

Other public outreach activities include a vast network of Web sites
and interactive visitor centers hosted by labs and university groups.
In addition to exploring the physics of the standard model,
accelerators and detectors, virtual visitors can run remote
simulations and control online detectors. These Web sites educate and
convey excitement about HEP, and visits often result in follow-up
questions which are sent to designated people.

Much of the outreach to Congress and the funding agencies is conducted
by people at the level of lab directors, who make frequent visits to
Washington, DC.  Similar visits by groups of physicists take place
annually or semi-annually.  These visits are most effective when they
are well-planned and scheduled shortly before relevant legislation is
debated and enacted.  The physicists need to have a clear idea of the
message they want to convey and be well-rehearsed and equipped to make
the desired impression.  According to people who have been on such
visits, a greater effect is achieved when young physicists take part
in the actual contact with members of Congress and staffers.  Another
form of outreach to Congress involves letter-writing campaigns, which
are advertised in the HEP community in times of need.  It is not known
how many physicists actually send in such letters.

Labs and universities maintain personal contacts with local media
representatives through public relations offices.  Interesting results
and important discoveries are released and communicated to the press
by these offices in a timely manner.  In addition, labs publish their
own periodicals which are sent to thousands of subscribers, including
some Congressional offices and physics teachers.

There are many teaching and mentoring programs that offer
opportunities to students and teachers of all levels.  For instance,
QuarkNet~\cite{QNet} is a flagship program of the National Science
Foundation (NSF) and the Department of Energy (DOE) that facilitates
the involvement of high-school teachers in HEP research, through
summer internships with active experiments.  Some 40 institutions
participate in the program, each receiving one year of funding plus
some lead-in and follow-up money.  Other activities include ``Saturday
Morning Physics'' lecture and demonstration programs for high-school
students, summer programs for undergraduates such as Research
Experiences for Undergraduates~\cite{REU} and internships for
minority students.  In addition, many individuals in national labs and
universities conduct their own outreach and education activities, with
very little or no financial support. The magnitude of such activities
is difficult to assess.

Many physicists recognize the importance of outreach and education,
and several programs facilitate outreach activities.  However, the HEP
funding levels for outreach are dismal when compared to those of NASA,
for example.  Due to legislation enacted in the mid-1990's, DOE can no
longer explicitly support pre-college education and outreach, resulting
in the cancellation of several K-12 programs.  The national
laboratories do provide some outreach funding from their general
operating funds at the level of 0.25-0.75\% of their operating
budgets, totaling about \$3 million. In comparison, NASA spends 2.2\%
of its operating budget on education and, in addition, individual
research grants are required to support outreach at the 1-2\% level.

NSF has no restrictions on education, and its support of education
programs throughout the sciences totals about \$600 million
annually. This support is not tied to NSF research grants. In
addition, Fermilab has a donor organization, the Friends of Fermilab,
that raises about \$300,000 per year.  It is much more difficult for
small university groups to obtain outreach support if they are not
part of an established program, such as QuarkNet.

\subsection{Results of the Town Meeting}
During the young physicists' town meeting at Snowmass, we polled the
approximately 100 attendees and asked for their opinions regarding
outreach and education.  The questions asked were mainly intended to
spark discussion, but ultimately generate either an affirmative or
negative response.  Several questions were intended to generate
discussion only.

\subsubsection{Role of Outreach}
\begin{figure}
\begin{center}
\epsfig{file=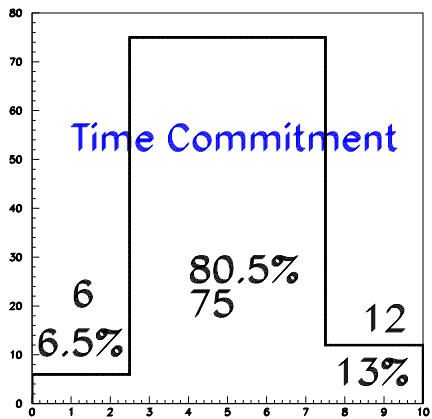, width=6.5cm}
\end{center}
\caption{A histogram displaying the number and percentage of young
  physicists willing to commit 1\%, 5\% or 10\% of their time to
  outreach.}
\label{fig:YPF_bloom_0717_fig1}
\end{figure}

The initial set of questions asked dealt with the role of outreach in
high-energy physics:

\begin{enumerate}
\item {\it Does HEP have a responsibility to do outreach? }
\item {\it Does HEP have an outreach problem?  Are we doing enough?}

There were 92 affirmative and one negative responses to these
questions.  The majority of young physicists felt that there is an
obligation to do outreach and that we are not doing enough.  Most felt
that outreach is not acknowledged as a part of our job and that
supervisors fail to regard time spent on outreach and education
activities as valuable.  One person commented that there is no reward
for those who do outreach and education.  Excellence in this capacity
should be acknowledged and rewarded.

A dissenting view was made regarding the claim that outreach will
generate more support for HEP, pointing out that a correlation between
outreach and increased funding levels had not been proven.

\item {\it How much time would you be willing to commit to outreach
    personally? (1\%, 5\%, 10\%?) }
  
The clear majority (80.5\%) of people attended were willing to
dedicate about 5\% of their time to outreach and education.  13.5\%
were willing to commit of order 10\% of their time, and 6.5\% were
willing to commit about 1\% of their time
(Figure~\ref{fig:YPF_bloom_0717_fig1}).  Those who abstained were not
explicitly counted in this vote, but we estimate the number to be less
than ten, based upon total head counts.
\end{enumerate}

\begin{figure}[ht]
\mbox{
\begin{minipage}{0.48\textwidth}
\centerline{\epsfxsize 3.0 truein \epsfbox{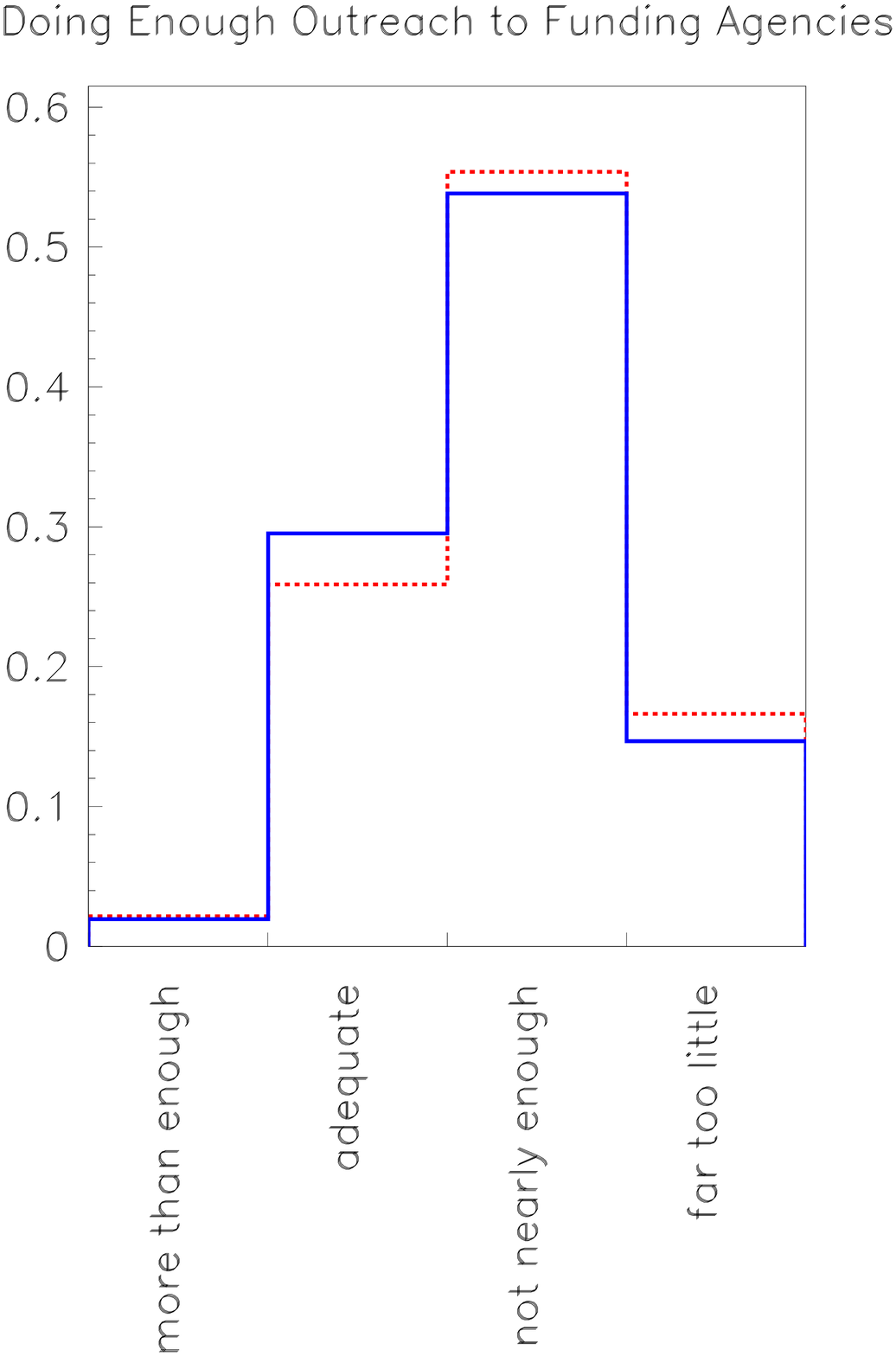}}
\vspace{-0.2in}
\caption{Are we currently doing enough outreach to funding agencies?
Solid lines are results
for those who identified themselves as young physicists, dashed lines
are for those who did not~\protect\cite{ref:ypppoll}.}
\label{fig:YPF_bloom_0717_fig2}
\end{minipage}\hspace*{0.02\textwidth}
\begin{minipage}{0.48\textwidth}
\centerline{\epsfxsize 3.0 truein \epsfbox{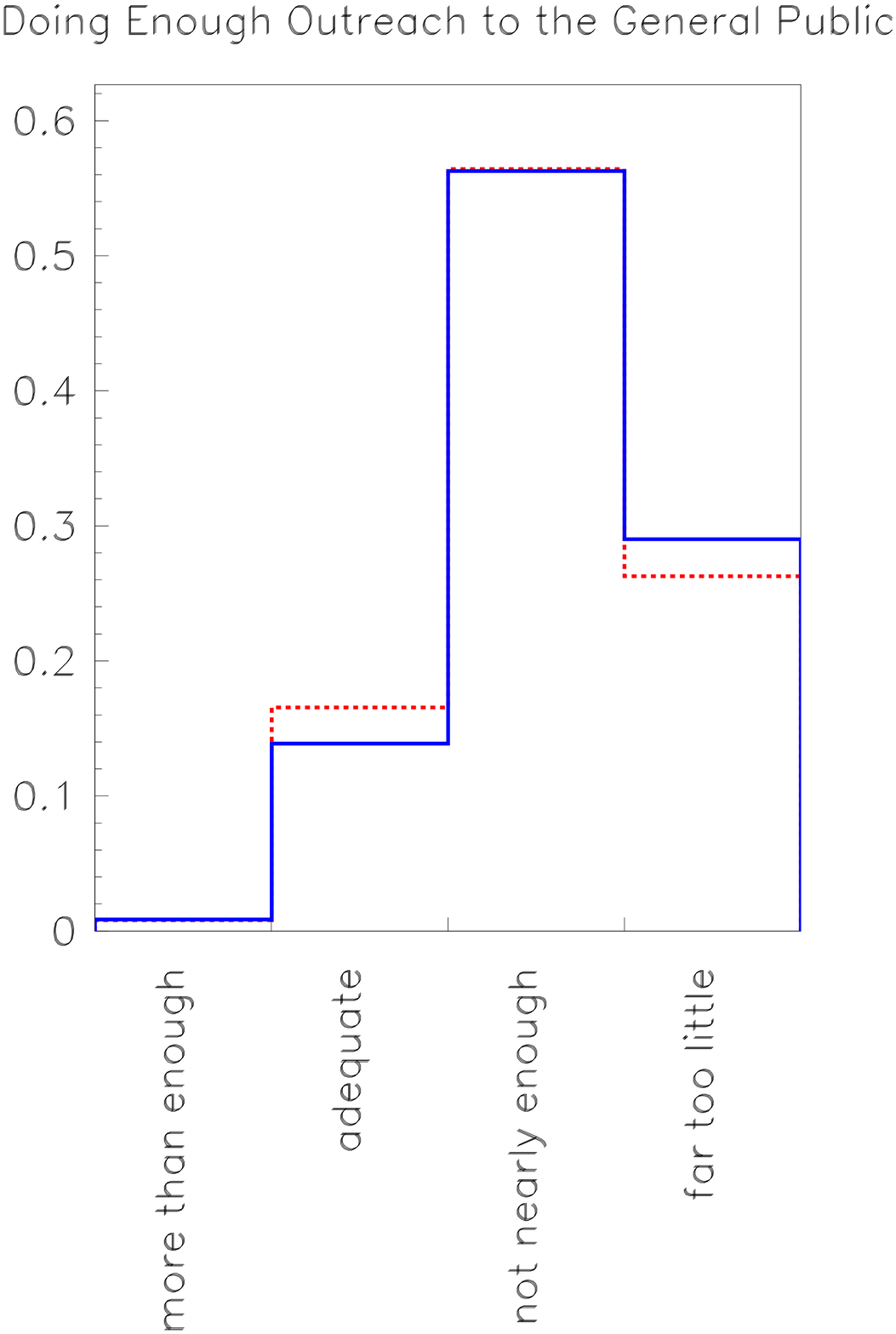}}   
\vspace{-0.2in}
\caption{Are we currently doing enough outreach to the general public?
Solid lines are results for those who identified themselves as young
physicists, dashed lines are for those who did
not~\protect\cite{ref:ypppoll}.}
\label{fig:YPF_bloom_0717_fig3}
\end{minipage}
}
\end{figure}

Similar results were found from the YPP survey~\cite{ref:ypppoll}. Two
questions were asked regarding outreach in high-energy physics:
(1)~Are we doing enough outreach to funding agencies
(Figure~\ref{fig:YPF_bloom_0717_fig2}) and (2)~Are we doing enough outreach to
the general public (Figure~\ref{fig:YPF_bloom_0717_fig3})?

\subsubsection{Tools and Mechanisms}

Outreach and education require a highly-organized effort in order to
effectively target and generate an excited audience. The following
questions were presented to the young physicists at Snowmass:

\begin{enumerate}
\item {\it Who should be in charge of organizing outreach?
    Physicists? Professionals?  Do we need an official organization
    (DPF) to coordinate outreach?}

The majority of responses to these questions were that we need
assistance in organizing the outreach activities.  The comments from
the audience were that we should arrange to have professionals
organize our activities, but that physicists should do the outreach
and have the actual contact with the target audience.

\item {\it Would you find outreach and education workshops
      useful?  Would you attend and contribute?}

This question provoked a positive response as well.  The majority said
that they would find such workshops useful and would attend and
contribute.  It was commented that such workshops should attempt to
unite physicists with science teachers so the groups can share ideas
and discuss pertinent issues.  Convenience is also a consideration. It
was pointed out that holding such workshops in advance of usual
physics gatherings would maximize attendance.

\item {\it Would you support grants or award incentives for
    people who perform outreach/education?   Should it be considered
    when making a new hire? }

The majority felt that dedication and excellence in outreach should be
rewarded and should be a consideration when a new position is filled.
\end{enumerate}

\subsubsection{Funding Issues}

The following questions were presented regarding funding of outreach
and education:

\begin{enumerate}
\item {\it Do you support raising HEP outreach funding
      to the level of NASA (2-4\%)?}

There were 88 people who supported raising the level to that of NASA
and 5 who were against doing so.

\item {\it Would you endorse DOE funding tagged specifically for
        outreach and education?  Should the DOE mission statement be 
        modified?}
      
A majority (89 in favor and 4 against) were in favor of modifying this
guideline to allow DOE to support line-item outreach and education
activities.

\item {\it If additional funds could not be located, what would you be
willing to sacrifice to promote education and outreach?  Any ideas for
additional support?}

This question was meant to spark comments from the audience. People
were in general unwilling to sacrifice physics activities for the
increased funding levels of HEP outreach.
\end{enumerate}

\subsection{Conclusion:  Our Commitment to Outreach and Education}
The young physicists at Snowmass have advocated spending 5\% of their
time, or approximately one day a month, on outreach and educational
efforts, in the belief that these activities are vital to the future
health of particle physics, and of physics and science in general.

To maintain momentum and the interest in outreach after Snowmass, we
young physicists are organizing our activities through the Young
Physicists' Outreach Program (YPOP). Our goal is to form seed groups
at labs and universities to discuss ideas and organize outreach
activities. To facilitate communication between the involved
individuals and groups, Web pages are being created, where people can
share their ideas and experiences, post materials and tools, and
maintain effort and interest levels. 

The ultimate goal is to build a network of people that will make it
easier for anyone to create or join an outreach project by building on
the resources and support of others.  Such a support network will make
it easier for physicists to get involved, despite their tight
schedules and other commitments.  Here are some specific things that
any physicist, individually or as part of a group, regardless of
affiliation, can do:

It is very important to open lines of communication with our
governments.  Congress should be an important focus, as it has the
final say on funding for our field, but it is also useful to teach
state and local governments about the benefits of physics to their
constituents.  Every university and national laboratory has an office
that is responsible for government relations.  Contact your government
liaison, and learn how you can communicate the importance of science
to our political leaders.  These liaisons will know who to talk to,
how to talk to them (what preparations you need to make for a visit,
whom and what to bring with you), and when to talk to them.  They can
help you establish relationships with people who are in a position to
make a difference.

Young physicists have typically made a very positive impression on
government leaders.  To make the most of this, a group of us are
currently working with Lewis-Burke Associates, a scientific-lobbying
firm, on a letter to Congress in support of high-energy physics.  This
letter will be circulated widely in our community for signatures.

In your local communities, you have many opportunities to interact
with the general public.  University-based researchers meet with some
members of the public every day -- our students.  By working with
lecture-demonstration staff, you can integrate particle-physics
demonstrations into introductory courses.  You can be a resource for
the local Society of Physics Students chapter, and promote the
opportunities of particle physics.  You can be a mentor to
undergraduates, or even high-school students, getting them into your
lab to see how research is done.  Many of us are in the field today
because a senior physicist reached out to us.

We are making efforts to contact science teachers and parents at all
levels in the public and private school systems. Our goal is to
provide a means for science teachers and parents to gain access to the
physicists who actually do the research.  Plans include making contact
with the American Association of Physics Teachers lead teachers,
learning from them what their needs are, bringing demonstrations and
new ideas into the classroom and working to increase science literacy
among parents.  We have begun to contact school principals in our
local areas and are working together with them on carrying out this
vision.  Emphasis is placed on active learning techniques to encourage
the development of science in the classroom.  YPP members are applying
to the NSF for funding to develop a set of instructional kits for
students in grades six through twelve, which would give them a
hands-on introduction to the tools and concepts of particle
physics~\cite{ref:yppweb}.

The NSF has expressed support for outreach to groups which are
traditionally under-represented in science. We are doing ourselves a
disservice if we do not get a wider range of people interested in, and
then involved in, scientific research.  We must cast our nets as
broadly as possible to bring in the brain power needed to understand
the physical world more thoroughly.  Programs focused on these groups
can make a difference for the field, and for the lives of those
involved.  But since members of under-represented groups can be found
just about anywhere, any activity that reaches a very broad spectrum
of the public will touch people who do not think about physics in
their daily lives. Any large event, such as a community festival or a
county fair, can be an opportunity to promote science to the larger
world.

To reach a much larger audience, you can get involved with print and
broadcast media.  Meet the staff at your local newspaper, and
encourage them to write about science, focusing on research at your
lab or university.  You can write opinion articles on physics or any
science-based topic that you are comfortable with.  Many local radio
stations, especially public stations, produce programs that sometimes
focus on science. You can volunteer to be a guest, or to steer them
toward interesting topics.  Some people are thinking really big,
producing ``particle commercials'' to appear on broadcast television,
and the APS helped the producers of ``The West Wing'' with a story
line about particle physics.  Such projects are very beneficial, but
you do not have to think that big to make a difference in your own
community; a simple letter to the editor gives a voice to science and
its practitioners.

The new research initiatives discussed at Snowmass will require
resources -- both financial and human -- beyond anything that our
field has known to date.  They will also require us to develop
stronger relationships with communities to host these new facilities.
If we want particle physics to have a robust long-term future, each
of us must take the time to reach out to the public and convince them
of the excitement and utility of our research. Even the simplest
effort counts toward creating the future.

\section{Globalization of High-Energy Physics}
\subsection{Introduction}
All current HEP facilities have risen by means of a single nation or 
region building and financing each project.  The resources required to 
build a new machine operating on the high-energy frontier appear to 
exceed what can be provided by one country.  Thus the goal of the  
Globalization Working Group (GWG) has been to examine issues associated 
with HEP projects on the global scale and how the U.S. HEP community 
might interact with the international HEP community to bring the next 
such project to fruition.  As young physicists, whose careers and 
research aspirations are tightly coupled to the development of the next 
major HEP facility, we believe that now is the critical time to review 
these issues.

The GWG has focused on the political and scientific structures that
can enable planning and funding of large experiments.  Particular
attention has been paid to the role of existing national laboratories
and preserving, or possibly improving, a researcher's ability to work
despite being located a large distance from an experiment.  These
questions were addressed during the young physicists' town meeting at
Snowmass 2001.  The specific topics discussed were international
steering, structure of an international laboratory, operation of an
international laboratory, and career issues.  Here we summarize those
discussions and present relevant supporting data from the YPP
survey~\cite{ref:ypppoll}.

\subsection{International Steering}
Although international collaboration has occurred for years, forming
an international consensus in HEP seems quite difficult. From time to
time, \textit{ad hoc} committees study the most pressing issues of HEP
and make recommendations to the community and funding agencies about
potential courses of research.  However, once their reports are
complete, such groups immediately disband.  Long-term planning and
continuity is very difficult with just these groups and there is no
guarantee that their findings will be widely accepted.  Similarly,
periodic gatherings of the community such as Snowmass workshops may
generate a degree of consensus, but the mechanisms for translating
such consensus into a plan of action, however, are unclear.  Permanent
groups such as the International Committee for Future Accelerators
(ICFA)~\cite{icfa} may not have the right structure and political
weight to serve as a steering group for the entire HEP community.

To this end, an international steering group, composed of members of
the international HEP community along with representatives from the
government funding agencies, might provide an effective mechanism to
support the development of energy-frontier facilities over the long
term.  A perceived advantage of such a group is that national funding
agencies may give more active support to long-range HEP efforts if
they are involved at this level.  In addition, it would provide a
forum for international coordination among the funding agencies, thus
making large projects easier to fund in a coherent manner.  It is even
conceivable that such a group might eventually manage the distribution
of funds for an international project.  On the other hand, there is
some concern that international coordination of a project may result
in a loss of national pride for the individual countries involved.
Governments may react adversely to not being in control of such a
large project.  Finally, the HEP budget for any given country might
decrease because other nations are contributing to the same project.

There were three straw poll questions asked to the young physicists at
the town meeting related to international steering. The first one was:
Should funding agencies rely more heavily on advisory bodies like
ICFA? About 50\% of the young physicists responded yes, and a few
responded no while the rest abstained from answering. The second
question was: Do you think an international committee should determine
the type/location of the next machine? Again about 50\% of the young
physicists responded yes, and a few responded no while the rest
abstained from answering. The third question was: Should an
international group eventually have funding power? In this case, the
majority of the young physicists abstained from answering, while a few
answered yes, and a few answered no. Apparently, more information is
needed about the mechanism of such a funding power in order to answer
that last question.  Indeed, the many abstentions in this section
probably indicate a lack of familiarity with the ideas of
international steering in general.  We encourage wider discussion of
these issues in the HEP community.

\subsection{Structure of an International Laboratory}
The funding and management structure of an international laboratory
can take several forms, and ICFA has considered the four basic models:

\begin{itemize}
\item National or regional facilities (``FNAL/DESY model''): the
facility is built and operated by a host country or region.  Planning
and project definition are handled by an international collaboration.

\item Larger facilities (``HERA model''): contributions to
construction are made by multiple countries.  Operation
responsibilities and costs belong to the host country.  Planning and
project definition are handled by an international collaboration.

\item  Very large shared projects:  all countries involved share the
responsibility and costs for construction, operation, and project definition.  
The facility is the {\em common property} of all. This option has never been 
attempted.

\item   Very large self-organized projects (``CERN model''):  the
participating countries do not directly manage the lab; rather, 
they contribute to an international organization which autonomously 
runs the projects.

\end{itemize}

When considering these models in the context of the next
energy-frontier project, a plurality of the town-meeting participants
(40\%) favored the solution of a very large shared project. The
remainder of the (non-abstaining) participants favored the following
either a self-organized structure (CERN model) (13\%-18\%), or a
national organization that receives contributions from international
groups (FNAL/DESY model)(13\%).

\subsection{Operation of an International Laboratory}
In the context of a very large shared project (see above), the
operation of an international laboratory offers special challenges.
In particular, since many collaborating institutions will contribute
to the facility's operation, key personnel resources will not be based
at the main laboratory.  In response to this issue, ICFA recently
appointed a working group to examine whether the implementation of a
Global Accelerator Network (GAN)~\cite{GAN:Wagner}, as proposed by
Albrecht Wagner (DESY), could provide a viable method for accelerator
physicists at widely-separated locations to support the operations of
the main laboratory.  At the heart of the GAN concept is the idea of
remote control rooms located at various regional centers around the
world for accelerator operation.  These centers would offer a fast,
reliable and secure network connection to the main laboratory for safe
remote control of the accelerator facility.  Advanced communications
capabilities would also be available for effective real-time
interaction among widely-scattered researchers, particularly when
diagnosing problems.  The preliminary conclusion of the ICFA working
group is that remote accelerator operation presents a number of
challenges but can be readily achieved~\cite{GAN:PAC2001}.

GWG discussions at Snowmass explored a more general concept of
regional centers as locations for remote detector control and data
analysis in addition to remote accelerator control.  An obvious choice
of location for such centers would be existing national labs, where
much of the infrastructure required for a regional center already
exists.  This location would also allow considerable interaction
between members of national experiments and the international project.
Smaller regional centers organized around universities might also be
feasible and would offer greater accessibility for the
widely-distributed members of the HEP community.  A number of benefits
could result from the regional-center concept. It would allow control
and responsibility to be distributed among countries other than the
host country.  The possibility of each regional center periodically
assuming primary operational responsibility would highlight national
contributions to the project and might particularly appeal to national
funding agencies.  Regional centers would also help prevent the
primary lab infrastructure from being overwhelmed with visiting
scientists and be more attractive to students because of easier
access.  Finally, they would help ensure the continued viability of
the HEP community in regions that do not actually host the next large
facility.  Potential drawbacks are the expense of duplicating the
infrastructure needed for control and analysis functions as well as
the speculative time frame for viability.

Results from the YPP survey and straw-poll questions posed at the town 
meeting provide an interesting window on young physicists' perspectives 
of how the next large international project might work.
In the event that the project is not located in the U.S., most 
physicists, both young and tenured, responded to the survey that the 
quality of beams and physics results would improve given the availability  
of regional centers in the U.S..  The survey also showed that most physicists 
found it desirable to have the same number of regional centers as existing 
national laboratories, consistent with the fact that the required 
infrastructure is most easily implemented at the national labs.  

At the town meeting, it was agreed by a 2-to-1 margin that the concept
of remote data-analysis facilities was viable.  This seems quite
consistent with ongoing efforts to provide remote data processing
capability for the LHC experiments.  When asked whether remote
accelerator operation was viable, the audience was evenly divided.
More work to demonstrate the validity this concept is clearly required
in order for it to be generally supported.  When asked whether remote
detector operation was possible, the audience answered ``no'' by a 2-to-1
margin (with approximately 60\% of those present casting votes).

\begin{figure}[ht]
\mbox{
\begin{minipage}{0.48\textwidth}
\centerline{\epsfxsize 3.0 truein \epsfbox{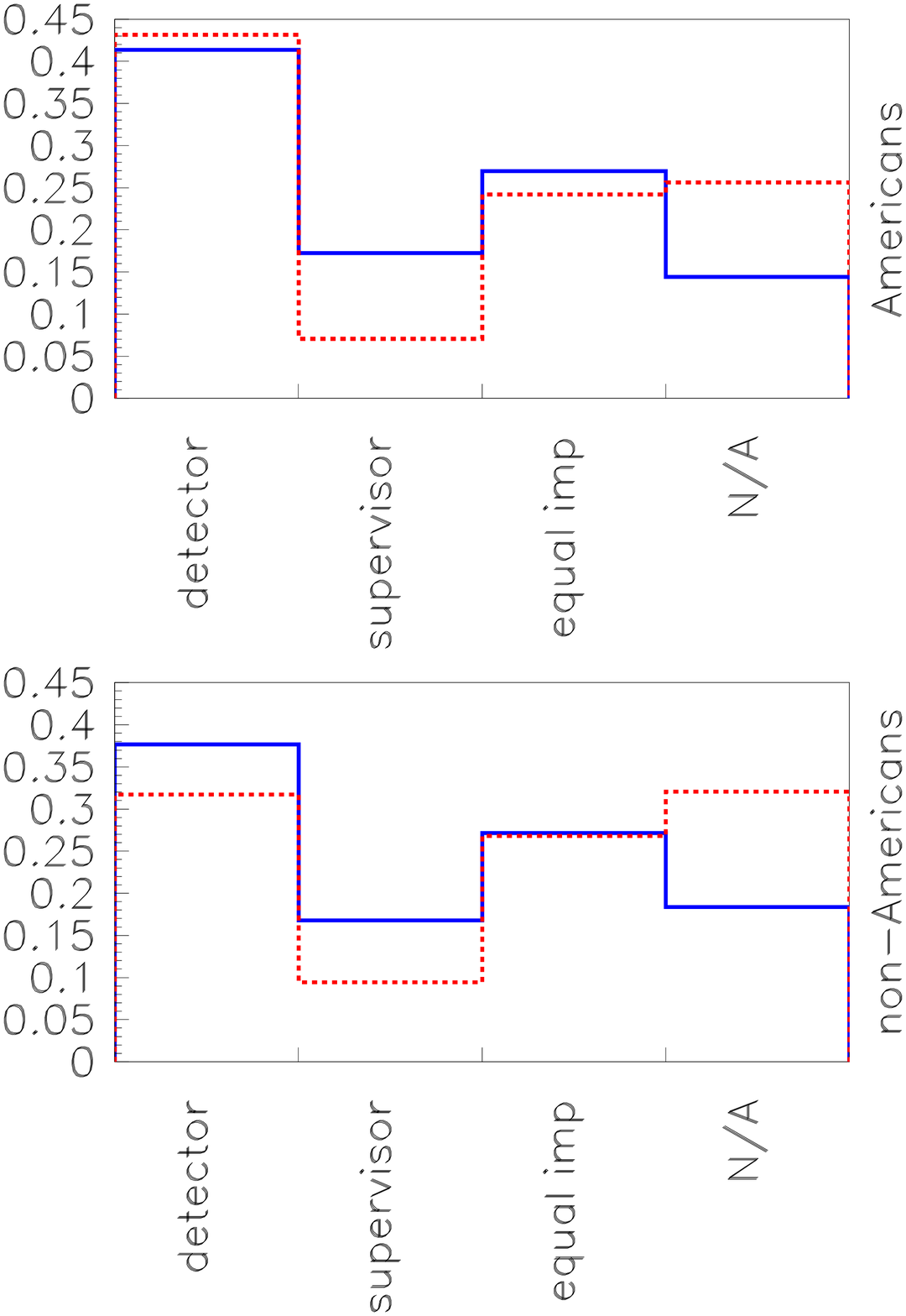}}
\vspace{-0.2in}
\caption{Given the choice, how do you rate being near your detector
versus being near your advisor/supervisor?  Solid lines are results
for those who identified themselves as young physicists, dashed lines
are for those who did not~\protect\cite{ref:ypppoll}.}
\label{fig:YPF_bloom_0717_fig4}
\end{minipage}\hspace*{0.02\textwidth}
\begin{minipage}{0.48\textwidth}
\centerline{\epsfxsize 3.0 truein \epsfbox{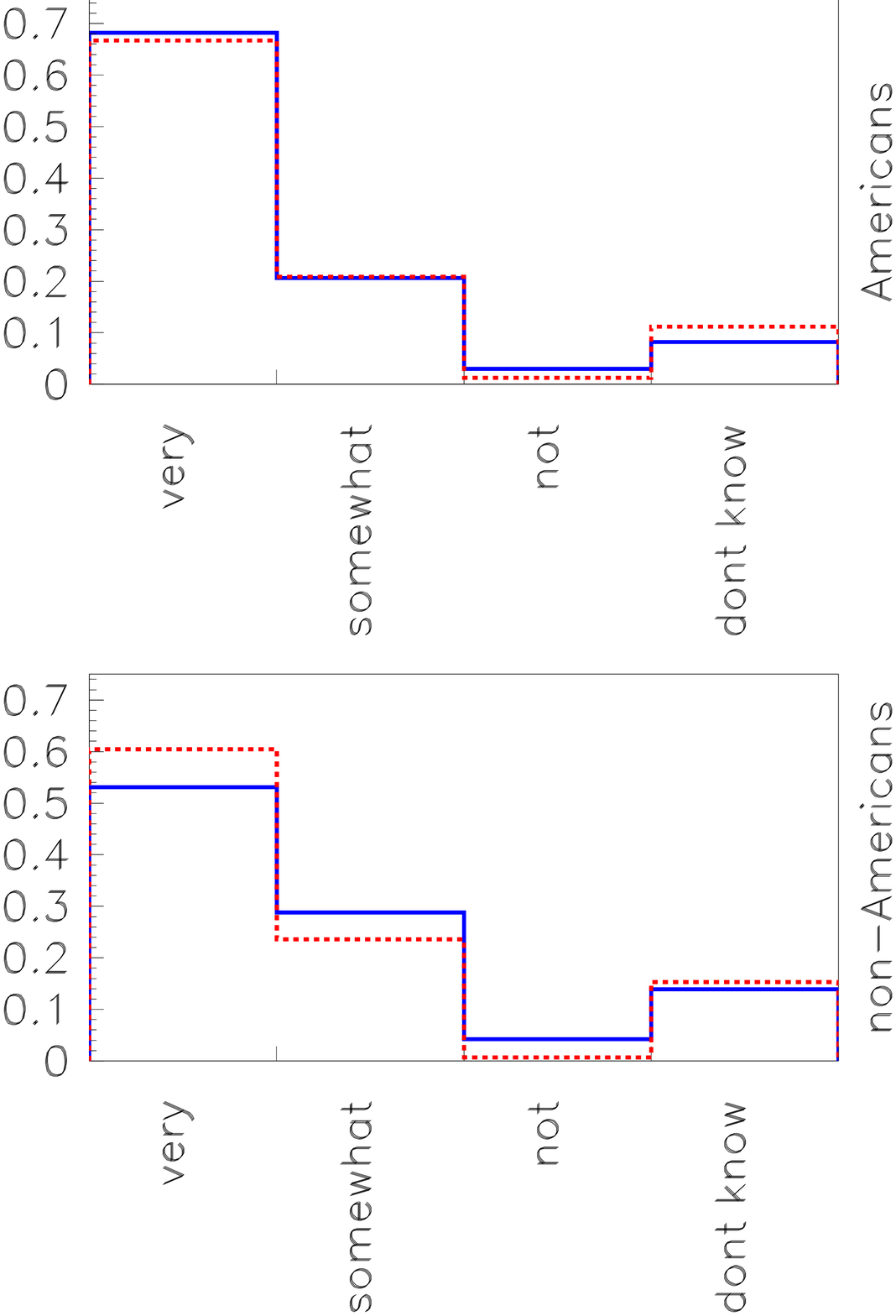}}   
\vspace{-0.2in}
\caption{How important do you feel it is to have hands-on hardware
experience?  Solid lines are results for those who identified
themselves as young physicists, dashed lines are for those who did
not~\protect\cite{ref:ypppoll}.}
\label{fig:YPF_bloom_0717_fig5}
\end{minipage}
}
\end{figure}

Greater insight into the issue of remote detector operation can be
obtained by looking at a survey question asking whether proximity to a
detector or one's advisor is more important for a graduate student.
The majority of physicists, across all categories sampled (young and
tenured, American and non-American), felt that proximity to the
detector is more important (Figure~\ref{fig:YPF_bloom_0717_fig4}).  Hands-on
experience and interaction with local experts are considered a key
piece of graduate education (Figure~\ref{fig:YPF_bloom_0717_fig5}).  Thus, the
training impact of remote versus local operation needs to be carefully
considered when exploring the possibility of remote operation of the
next large facility.  All in all, we strongly encourage continued
study of GAN models and how they might be adapted to a future facility
as well as detailed tests of remote control from sites removed from
the physical laboratory.

\subsection{Career Issues}
A final topic of particular relevance to young physicists in an era
where large international collaborations are the norm is that of
careers which often cross international borders.  At the town meeting,
a number of young physicists expressed the opinion that Europeans
typically have an easier time getting permanent jobs in the U.S. than
Americans do in Europe.  European Union rules with respect to
obtaining university positions were a point of particular concern.  We
believe that the HEP community needs to understand and review such
policies and work to enhance the ability of young physicists to pursue
careers within international collaborations without regard to the
borders separating the various institutions. In the case of
U.S. institutions, it may also be a good idea to revive the
second-language requirement as part of the graduate physics curriculum
to improve job opportunities for native English speakers in
non-English speaking countries.

\subsection{Conclusions}
From the town-meeting experience, it is clear that young physicists
are relatively unaware of the international groups currently studying
the future of our field and how they operate. Wider distribution of
information about such groups is the first step to ensuring greater
participation of the community.  It will also allow the creation or
modification of such groups as the need arises.

Young physicists are more certain about what kind of organization
would be appropriate for the structure of the next international
project. It is interesting to note that young physicists are well
aware of the challenges ahead, and seem to favor a completely new type
of organization to face those challenges. In terms of the operation of
this future laboratory, there was more dissension among the
participants, reflecting the same situation among the entire HEP
community. What is clear for young physicists is that interacting with
the detectors and accelerator, participating in the crucial decisions
and exchanging ideas with other physicists about the physics being
done is of utmost importance. Hence, the use and structure of regional
centers will have to carefully address these fundamental requirements.

The underlying thread of globalization issues in HEP reflects similar
trends in our social structures: in order to achieve our physics
dreams, we have come to a point where all members of the high-energy
physics community, around the world, are needed. Achieving the
highest-quality work is a major challenge in these
circumstances. There is no doubt that the success of this enterprise
will be a motivating example for the rest of society, as science
should be.

\section{Balancing and Building the Field}
\subsection{Introduction}
The overall goal of the Balancing and Building the Field Working Group
is to understand the broad physics interests of the young members of
the high-energy physics community, and what specifically the field
should do to balance these interests.  The working group addressed
four main questions facing the field today: How should we define the
field of high-energy physics for the next 20 years?  What are the
physics questions we should address, and how should we prioritize
them?  What future facilities are needed to address our most important
physics goals?  Do we have the necessary number and type of scientists
to accomplish our goals?  We provide a summary of some of the answers
to these questions here.

\subsection{Defining the Field}

As collider-based programs move to higher energies, the cost and
number of people required seem to continually increase.  The
recognition of this trend prompts us to question whether this is
sustainable.  Will there be enough funding or enough people to support
more than one or two large experiments in the future?  Are we limiting
the scope of high-energy physics by referring to it as ``the energy
frontier?''  Do we need to broaden our perspective to be particle
physicists, not just high-energy physicists?

Nuclear physics today is often called ``medium-energy physics''
because it operates at energies that were solely the domain of HEP only a
few decades ago.  These regimes of QCD remain extremely important to the
HEP programs.  The young community has expressed a great concern that,
despite the common interests, there is not enough interaction between
nuclear and particle physicists.  Some of the comments form the town
meeting were:
\begin{quote}
\textit{At the least, for the sake of the science, cross talk between
the particle and nuclear fields needs to improve.}

\textit{Perhaps there should be a more tangible joining of the fields.}

\textit{Integration of the funding agencies could promote cross talk $\ldots$
But perhaps it could also negatively impact overall funding levels?}
\end{quote}

In the United States, there are two successful nuclear programs
currently in operation that were funded at the level of \$1 billion
each: the Relativistic Heavy Ion Collider at Brookhaven National
Laboratory, and Continuous Electron Beam Accelerator Facility at
Jefferson Laboratory.  The funding for these facilities is considered
independent of the funding for particle physics.  The DOE has separate
particle and nuclear physics divisions that appear to have very little
cross-communication.  This is in contrast to Europe, where many
programs are connected, {\it e.g.} the LHC will convert to running
heavy ions at some point.

Addressing the concerns of balancing scientific goals with funding
realities, we asked ``Should HEP and nuclear physics join funding
efforts?''  The overwhelming consensus was ``no!''  While there is a
desire for closer scientific ties to nuclear studies, the programs,
while complementary, set out to ask fundamentally different questions.

Given the strong response against joining funding efforts, we also
asked ``Should we create a formal mechanism to combine HEP with
related physics programs, but retain independent funding sources?''
The idea was to set up an agency that could facilitate the
cross-disciplinary exchange of ideas and data.  While there was
interest in such a body, the conclusion was that more specific
proposals need to be considered.

\subsection{High-Energy Physics and New Facilities}

Both the YPP survey \cite{ref:ypppoll} and the Forum results indicate
that young physicists believe that a major new high-energy physics
facility is needed in order to elucidate the most pressing current
problems in particle physics.  More than 80\% of YPP survey
respondents who identified themselves as young answered either
``yes'' or ``maybe'' to the question.

From the survey results, an overwhelming majority of young physicists
consider the understanding mechanism of electroweak-symmetry breaking
(EWSB) to be the most important of a large number of subjects to the
field.  Since the potential for discovering and studying the EWSB
mechanism is greater at new, higher-energy facilities, this focus of
interest almost certainly explains why so many young physicists
believe that a new facility is necessary.

Because physics potential must determine which facilities will be
built, many believe that the HEP community should be certain that it
has sufficient knowledge to proceed with designing and building a
major new facility.  At the town meeting, the young physicists in
attendance were asked if now is the time to make a decision on the
next facility.  More than half answered ``decide now'' while the
remainder were split between ``abstain'' and ``wait.''  Those
answering ``wait'' were further asked if the reasons for waiting were
technologically based, physics based, or both.  About half said
``physics'' while the other half said either ``technological'' or
``both.'' Some comments (paraphrased here) preceding the vote included
the following:
\begin{quote}
\textit{We will lose accelerator expertise if we wait.}

\textit{The unexpected brings discovery...decide now!}

\textit{We should wait to see what the LHC sees!}

\textit{What if we are looking in the wrong place?}

\textit{Does the next facility have to be accelerator-based?}
\end{quote}

But it was also clear that smaller experiments should not be
sacrificed in order to obtain this facility. It was felt that a broad
and diverse program of physics ought to be maintained both for the
health of the field and because many young physicists are interested
in physics that cannot be done at a large energy-frontier
facility. Indeed, the survey revealed that a significant number of
survey respondents find QCD, the neutrino sector, exotic searches, or
cosmology to be most personally compelling.

However, EWSB at a major new facility seems to be the clear priority
among young physicists. Given that both the Tevatron and the LHC are
either already engaged, or soon will be, in the search for the
mechanism of EWSB, the question then is what sort of facility is
required beyond these hadron colliders.

Table~\ref{tab:YPF_bloom_0717_tab1} summarizes the options currently under
consideration. Sources include talks addressed to the 2001 HEPAP
long-range-planning subpanel~\cite{HEPAPWWW}, various HEPAP
reports~\cite{HEPAPREP}, the TESLA Technical Design
Report~\cite{TESLATDR}, and the Linear Collider Physics Resource Book
for Snowmass 2001~\cite{LCRES}.  Further informational resources can
be found at New Approaches to a Very Large Hadron
Collider~\cite{VLHCRES} and the Neutrino Factory and Muon Collider
Collaboration~\cite{MUCOL}.  All estimates have been taken from the
documents and resources listed above, but they are necessarily
provisional and subject to debate. In particular, accounting systems
differ and some estimates may not include labor, detectors,
contingency and other costs.
\begin{table}[ht]
\begin{center}
\caption{Major new HEP facilities under consideration. Cost 
estimates are necessarily provisional and accounting systems differ.}
\bigskip
\begin{tabular}{|l|c|c|c|c|} \hline
    Facility & Decision Point & Construction Time & Turn-On & Cost \\
    \hline
    TESLA & now & 8 years & 2010? & 3.1B eur $\approx$ \$2.7B \\
    NLC/JLC & 2003 & ? & ? & \$6B \\
    $p$  Driver & 2004 & 4 years &  2008-2010? & \$0.3B \\
    $\nu$ Factory & 2006 & 6 years & 2013? & \$1.9B \\
    $\mu$ Collider & 2010-2015 & 8 years & 2018-2025? & ? \\
    VLHC & 2010-2015 & 10 years & 2020? & \$4B-\$5B \\
     \hline
\end{tabular}
\label{tab:YPF_bloom_0717_tab1}
\end{center}
\end{table}

At the town meeting, the young physicists were asked: ``What future
flagship facility should we build?'' The voting options were linear
collider, VLHC, muon collider/neutrino factory, or abstain.  The votes
were 35 (51\%) for the linear collider, 5 (7\%) for the VLHC, 19
(28\%) for the muon collider/neutrino factory with 10 (14\%)
abstentions.  It was further asked if the next machine had to be in
the U.S. The answer was overwhelmingly ``no,'' and when asked if the
U.S. should join efforts if the next machine was built outside the
U.S., the response was overwhelmingly ``yes.''  Some comments
preceding these votes were:

\begin{quote}

\textit{We should take the path that gets us the best physics on
the shortest time scale --- regardless of location.}

\textit{A neutrino factory has the advantage of being a stageable facility.}

\textit{The most realistic option on the list is TESLA.}

\end{quote}

\begin{figure}[ht]
\centerline{\epsfxsize 3.0 truein \epsfbox{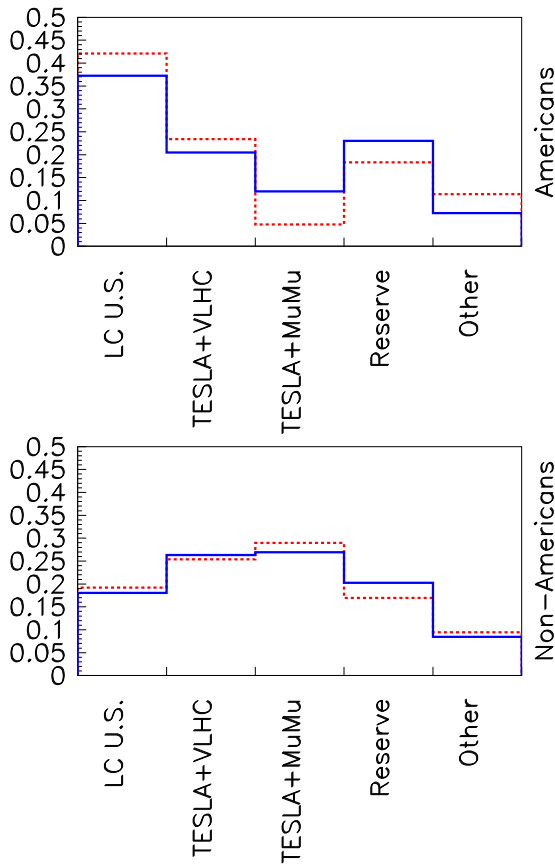}}
\caption{Responses to the YPP survey question on preferred new
facilities.  Solid lines are results for those who identified
themselves as young physicists, dashed lines are for those who did
not~\protect\cite{ref:ypppoll}.}
\label{fig:YPF_bloom_0717_fig6}
\end{figure}

The YPP survey asked: ``If you had to choose from the following
options, which one would you pick?''  The options included a new
$e^+e^-$ linear collider in the U.S. with continued research in VLHC
and muon ring/collider technologies, TESLA in Germany soon with the
goal of a VLHC in the U.S., TESLA in Germany soon with the goal of a
muon ring/collider in the U.S., reserve judgment for several years and
continue research, or other.  The results are shown in
Figure~\ref{fig:YPF_bloom_0717_fig6}. Young American physicicts are
almost evenly split between a linear collider in the U.S., and TESLA
in Germany with an eventual muon ring/collider or VLHC in the U.S. The
U.S. linear collider option earned a few more votes.

The conclusions to be drawn from the Forum and from the survey are clear. 
Young physicists believe that HEP needs a linear collider to complement
the search for the mechanism of EWSB at the Tevatron and LHC, but that
it should not be obtained by sacrificing smaller experiments and other
interesting subjects in the field. While no consensus on the location of the
linear collider has emerged, the majority of young physicists at the 
town meeting believe that the time to decide is now.

\subsection{Maintaining the Number of Physicists We Need}
The American Institute of Physics has kept statistics on the number of
entering physics graduate students for several years.  In a recent
report \cite{AIP:survey01} it was shown that the number of graduate
students from the United States has declined by nearly a factor of two
in the past ten years.  State universities have had to lobby to change
regulations governing the permitted number of foreign graduate
students in order to slow the trend of declining enrollments.  Even at
current levels it is not clear that there will be enough people in the
future to accomplish the physics goals of the field.

\begin{figure}[ht]
\mbox{
\begin{minipage}{0.48\textwidth}
\centerline{\epsfxsize 3.0 truein \epsfbox{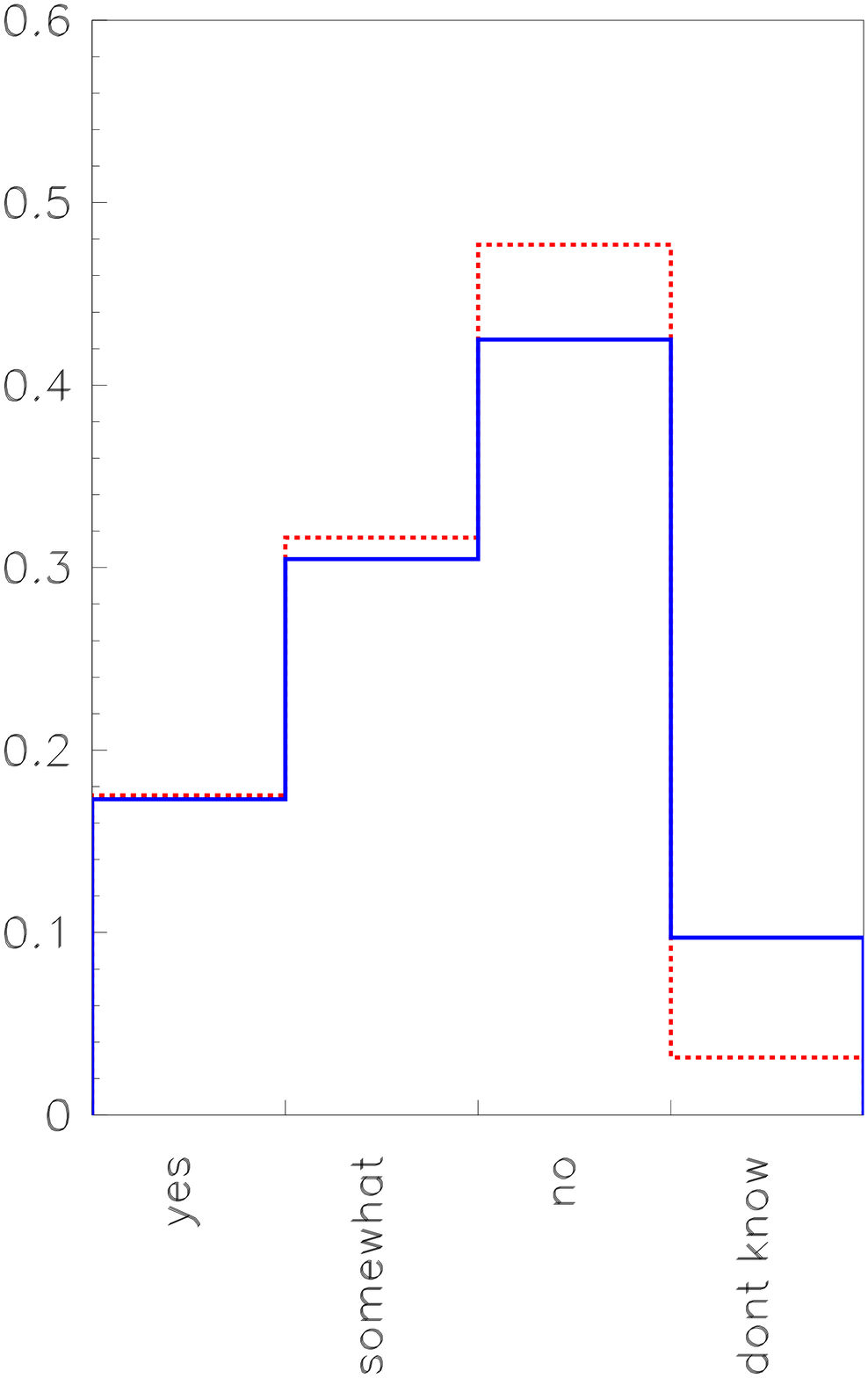}}
\vspace{-0.2in}
\caption[Are we retaining adequate numbers of talented physicists in
HEP?]  {Do you think we are currently retaining adequate numbers of
talented physicists in HEP?  Solid lines are results for those who
identified themselves as young physicists, dashed lines are for those
who did not~\protect\cite{ref:ypppoll}.}
\label{fig:YPF_bloom_0717_fig7}
\end{minipage}\hspace*{0.02\textwidth}
\begin{minipage}{0.48\textwidth}
\centerline{\epsfxsize 3.0 truein \epsfbox{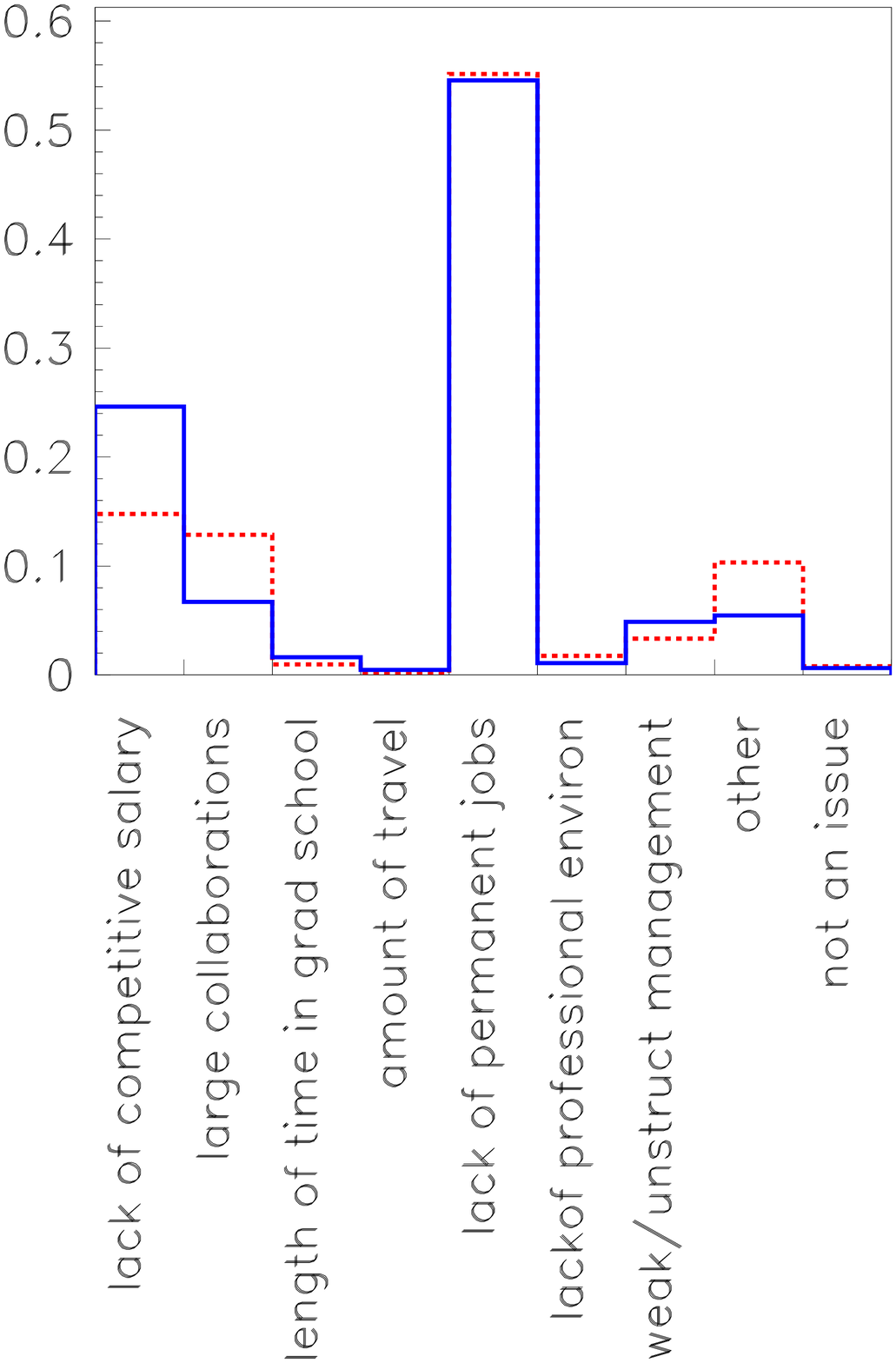}}   
\vspace{-0.2in}
\caption[What most influences young people to leave the field?]  {What
feature of the field do you think might \textbf{most} influence young
people to leave the field? Solid lines are results for those who
identified themselves as young physicists, dashed lines are for those
who did not~\protect\cite{ref:ypppoll}.}
\label{fig:YPF_bloom_0717_fig8}
\end{minipage}
}
\end{figure}

During the town meeting we asked whether we were retaining enough high
energy physicists, and the overwhelming consensus was ``no.''  As
shown in Figure~\ref{fig:YPF_bloom_0717_fig7}, less than 20\% of
respondents to the YPP survey felt that high-energy physics was
retaining enough physicists.  The survey asked ``What feature of the
field do you think might \textbf{most} influence young physicists to
leave the field?''  The responses are shown in
Figure~\ref{fig:YPF_bloom_0717_fig8}.  The majority of people are
leaving because they simply cannot find a permanent position.  The
other significant cause is a lack of competitive salaries.

Two areas of immediate concern were identified during Snowmass: the loss
of experimentalists who do more research and development work (R\&D)
than analysis and the loss of phenomenologists.  The attrition rate
for both of these groups was considered so advanced that the field
already may be in trouble.

At the town meeting, the Advanced Detector Research Project (ADRP) was
cited as a good example of what we can do to support detector R\&D.
The ADRP was begun by the Department of Energy in 2000 with the
purpose of ``supporting the development of the new detector
technologies needed to perform future high-energy physics
experiments.''  So far six grants have been awarded totaling
approximately \$500,000, with a desire to increase the funding to over
\$1,000,000 over the next few years.  Support for an increase in the
number of competitive grants offered to experimentalists who do R\&D
would send a positive message to departments that these
experimentalists are the backbone of the high-energy physics
community.

During the town meeting the question was asked: ``Should we dedicate
additional permanent positions to experimentalists who do primarily
R\&D work?''  Nearly 40\% of the audience agreed this was important.

Do we have the right balance of experimentalists, phenomenologists,
and formal theorists?  The clear consensus at the town meeting was
``no!''  Concerns were expressed that the field is not supporting the
people who make connections between the experiments that are running
and the theories we wish to test, while allotting too much of our
resources to string theory.  Young theorists who have developed much
of the current phenomenological technology are being forced to leave
the field because they cannot find permanent positions.

In order to judge the depth of the problem, we asked ``If you had one
permanent position to give away, would you mandate that it went to a
phenomenologist?''  A surprising 1/3 responded ``yes'' with most of the
rest uncertain.
Given that over 70\% of the audience said either that the next
position had to go to a phenomenologist or experimentalist
concentrating on R\&D, it is clear that these two groups should be
given a priority in hiring until the situation is corrected.

\subsection{Conclusions}
The Balancing and Building the Field Working Group has provided a
forum for young physicists to express their perspective on where the
field is and where it should be going.  We recognize that, while the
future of the field will include physics ``at the energy frontier,''
we will also have to leverage the understanding gained in nuclear and
astrophysics experiments.

The most important physics questions drive the decisions about what
facilities we will need in the future.  The majority of young particle
physicists believe that electroweak-symmetry breaking is the most
important phenomenon to understand.  We believe that a linear $e^+e^-$
collider is an essential element in achieving this understanding, and
is necessary for the future of our field.  However, one machine cannot
solve all of the important problems.  We must maintain a diversity of
experiments.

The working group believes there is a serious personnel problem in
high-energy physics.  Too many of our essential younger physicists are
having to leave the field because there are no permanent positions for
them.  In particular, we are losing too many experimentalists who
concentrate on R\&D and theorists who do phenomenological research.  A
concerted effort should be made to provide more permanent positions
for these talented individuals.

The future of high-energy physics is very promising.  The next 30
years will hold many surprises, discoveries, and difficult challenges.
Young physicists are prepared to lead the field to success.

\section{Perspectives}
In general, the opinions and aspirations of young physicists are not
that different from those of older physicists.  We are all interested
in keeping the field healthy and lively over the coming decades, and
have similar ideas on what actions are needed to do so.  But the Forum
did bring out several interesting points.  Young physicists expressed
a clear interest in increasing the funding and level of effort for
outreach and educational activities, and are willing to put up to 5\%
of their own time into it, despite concerns about the reward system.
They believe that globalized operations through the Global Accelerator
Network and regional centers can work, but there are still many
questions about the details.  They are looking for creative solutions
to the ``structural'' problems of the field, such as the lack of
permanent positions, especially for young phenomenologists and
detector builders; the possibility that we do not have enough people
to achieve our research aspirations; and concerns about the effects of
long lead times for experiments.  The opinions of young physicists on
future experiments and facilities seem to fit in with the mainstream.
There is a desire for a balanced program, including astrophysics, with
electroweak-symmetry breaking studies as a top priority.  An $e^+e^-$
linear collider, built somewhere, somehow, is the preferred next
machine.  While there is a slight preference among U.S.-based
physicists to have the world's next major facility built in the U.S.,
it is by no means a requirement.

The Forum had a positive tone, with much constructive discussion and
thinking about the long-term future, and we were very pleased by the
community's reaction to our efforts.  The Snowmass workshop gave young
physicists a chance to get to know each other and build the trust that
is necessary for the future of the field.  We believe that young
physicists are eager to make the case and do the work to understand
the fundamental laws of nature, and that our creativity and enthusiasm
will keep HEP healthy for a long time.  For instance, young physicists
were full participants in the Snowmass working groups, and helped
produce many of the results in these proceedings.  We cannot know yet
if we have come up with the ``right'' answers to the questions facing
us, but we are trying to work them out for ourselves.

Young physicists plan to remain engaged in the community.  The YPP,
which has been renamed the Young Particle Physicists, has reorganized,
and is forming new chapters at various universities and labs, such as
SLAC and DESY, to work on issues raised for the Forum at the
grass-roots level.  While the Young Physicists' Forum ended with the
Snowmass workshop, we will carry on future projects with the ``spirit
of Snowmass'' -- a spirit of openness and cooperation as we work to
create the long-term future of the field.

\begin{acknowledgments}
We thank the Organizing Committee for their sponsorship of the Forum,
and their efforts to promote the interests of young physicists.  Chris
Quigg in particular was unstinting in his support of our work.  We
also thank the APS, DPF, and DPB, and their respective officers, and
the NSF for their financial support of young physicists at Snowmass,
through the registration-fee waiver and low-cost housing for graduate
students and the Melvyn Month and DPF Snowmass Fellowships.  Cynthia
Sazama, Jeff Appel and the workshop staff helped us with many
logistical arrangements.  We are very grateful to the young physicists
who participated in the Forum activities, and to the community as a
whole for their positive reception of our work.
\end{acknowledgments}

\end{document}